\documentclass[runningheads]{llncs}
\usepackage{graphicx}
\usepackage{booktabs}
\usepackage{multirow}
\usepackage{siunitx}
\usepackage{grffile}
\usepackage{subcaption}
\usepackage{capt-of}% or \usepackage{caption}
\usepackage{bm}
\usepackage{amssymb}
\usepackage{amsmath}
\usepackage{hyperref}
\usepackage{isomath}
\usepackage{paralist}
\usepackage{tabularx}
\begin{document}
\title{Automated quality assessment using appearance-based simulations and hippocampus segmentation on Low-field paediatric brain MR images}
\titlerunning{Quality assessment and hippocampus segmentation}
\author{Vaanathi Sundaresan\inst{1,3} \and Nicola K Dinsdale\inst{2}}
%\orcidID{0000-0002-9451-4779}
\authorrunning{ Sundaresan et al. }
\institute{Department of Computational and Data Sciences, Indian Institute of Science, Bangalore 560012, India \and Oxford Machine Learning in NeuroImaging Lab, Department of Computer Science, University of Oxford, Oxford, United Kingdom \and Corresponding~author: 
\email{vaanathi@iisc.ac.in}
}
\maketitle              % typeset the header of the contribution
\begin{abstract}
Understanding the structural growth of paediatric brains is a key step in the identification of various neuro-developmental disorders. However, our knowledge is limited by many factors, including the lack of automated image analysis tools, especially in Low and Middle Income Countries from the lack of high field MR images available. Low-field systems are being increasingly explored in these countries, and, therefore, there is a need to develop automated image analysis tools for these images. In this work, as a preliminary step, we consider two tasks: 1) automated quality assurance and 2) hippocampal segmentation, where we compare multiple approaches. For the automated quality assurance task a DenseNet combined with appearance-based transformations for synthesising artefacts produced the best performance, with a weighted accuracy of 82.3\%. For the segmentation task, registration of an average atlas performed the best, with a final Dice score of 0.61. Our results show that although the images can provide understanding of large scale pathologies and gross scale anatomical development, there still remain barriers for their use for more granular analyses. 

\end{abstract}
\section{Introduction}

 Understanding the healthy development of structures within the paediatric brain is vital for the identification of neuro-developmental disorders. 
So far, application of image analysis tools on large scale MR imaging studies (e.g., the UK Biobank \cite{Biobank}) for automated analysis of healthy structural development of the adult brain has been well described. However, the majority of existing image analysis tools struggle with paediatric brains due to the poor gray matter/white matter differentiation and rapid growth and change observed in brain anatomical structures. This, combined with the challenges of imaging children, such as the need to minimise the movement of infants, critically limits our ability to understand the structural development of the paediatric brain. 

This knowledge gap is further increased when considering populations from Low and Middle Income Countries (LMICs) where high field MRI systems (1.5T / 3T) are rare due to the costs involved. Therefore, few MR imaging studies have considered LMIC populations. To fill this gap, Hyperfine SWOOP scanners are being tested - although at 0.064T the scanners offer much lower image quality, low-field MRI offers portability, cost-effectiveness and removes the need to sedate children during the scan. This improves the potential to increase the availability of MR imaging in underrepresented research communities \cite{LISA2024}. 

Existing image analysis tools (e.g., FSL, Freesurfer) have been primarily developed for 1.5 and 3T MR images of adults: the large domain shift between these images and the low-field paediatric images we are considering mean that the existing tools are unlikely to give high quality results \cite{Dinsdale2021}. Therefore, there is a need to develop domain specific tools, designed specifically for the low-field images and paediatric populations. Deep learning (DL) based methods are generally good candidates for producing analysis tools, such as segmentation methods, for low-field paediatric images, due to their ability to identify underlying discriminative patterns in images and create both local- and global-level associations between voxel values \cite{lecun2015deep}. Accurate segmentations of subcortical brain structures are essential for volumetric and morphological assessment and the monitoring of healthy brain development, but manual segmentation is time-consuming and error-prone \cite{balboni2022}. The hippocampus is a grey matter structure located within the medial temporal lobe memory circuit, and is associated with numerous conditions including depression,  Alzheimer’s disease, psychosis and epilepsy \cite{Anand2012}. Therefore, accurate delineation of the hippocampus is essential to enable volumetric and morphological assessment for understanding healthy development and disease.   Therefore, to assess the feasibility of low-field paediatric MR brain imaging, two initial tasks are considered:
\begin{itemize}
    \item Task 1: Automated quality assurance (QA) to rate the overall quality of low-field MRI images, to ensure that the acquired MR images meet specific standards.
    \item Task 2: Automatic segmentation of the bilateral hippocampi, due to their importance as a subcortical structure linked to cognitive and memory functions.  
\end{itemize}

Here we explore preliminary results from the above two tasks based on T2 weighted MR images by comparing multiple approaches. 
Our results show that these images with single modality can provide understanding of overall anatomies and large scale pathologies. However, our results suggest that multiple scan types and quality enhancement approaches might be required for more granular analyses and in-depth exploration of structural development in paediatric populations.

\section{Methods}

\subsection{Classification of artefacts for quality assessment}
We aim to perform quality assessment of paediatric brain MR images by assigning scores of 0, 1 and 2 (0 being good quality and 2 meaning there is a very high level of artefacts) across seven artefact domains such as noise, zipper, positioning, banding, motion, contrast and distortion. One of the major challenges is the imbalance in data samples across the different scores: the majority of cases were of good quality and the proportion of data with artefacts was extremely small, especially the ones with a score of 2. Hence, as a first step, we performed appearance-informed transformations to simulate artefacts to augment the data (for classes 1 and 2) for training. 

\begin{table*}[h!]
\centering
\scriptsize
\caption{Appearance-based transformation used for data augmentation (the parameters were selected empirically from held-out data).}
{\renewcommand{\arraystretch}{1.8}%
\begin{tabular}{l|c|c}%|>{\setlength{\baselineskip}{1.5\baselineskip}}c|}
  \hline
  \parbox[c][][t]{2cm}{artefact domain}& 
  \parbox[c][][t]{3cm}{Simulation steps}& 
  \parbox[c][][t]{7cm}{Parameters for steps}  \\
  \hline
  \parbox[c][][t]{2cm}{Noise (e.g. low-field strength)}& 
  \parbox[c][][t]{3cm}{Gaussian blur + Blur}& 
  \parbox[c][][t]{7cm}{\vspace{0.5em} \textbf{Class 1:} \\ $\sigma_{noise}$ = [0, 0.1], $\sigma_{blur}$ = [0, 0.6] \\ \textbf{Class 2:} \\  $\sigma_{noise}$ = [0, 0.2], $\sigma_{blur}$ = [0, 0.6] \vspace{0.5em}} \\
  \hline
  \parbox[c][][t]{2cm}{Zipper (e.g. EM spikes in gradient coil \& fluctuating power supply)}& 
  \parbox[c][][t]{3cm}{Herringbone artefacts + Blur}& 
  \parbox[c][][t]{7cm}{\vspace{0.5em} \textbf{Class 1:} \\ No. of spikes = 1, Intensity contrast (between spikes) = [0.1, 0.3], $\sigma_{blur}$ = [0, 0.6] \\ \textbf{Class 2:} \\  No. of spikes = 1, Intensity contrast (between spikes) = [0.3, 0.6], $\sigma_{blur}$ = [0, 0.6], $\sigma_{blur}$ = [0, 0.6] \vspace{0.5em}} \\
  \hline
  \parbox[c][][t]{2cm}{Positioning (e.g. discrepancy and phase- and frequency-encoding sampling times)}& 
  \parbox[c][][t]{3cm}{Translation + Rotation + Blur}& 
  \parbox[c][][t]{7cm}{\vspace{0.5em} \textbf{Class 1:} \\ $x_{offset} \in$ [-10, 10] voxels, $y_{offset} \in$ [-10, 10] voxels, $\theta \in$ [0, 10] deg, $\sigma_{blur}$ = [0, 0.6] \\ \textbf{Class 2:} \\  $x_{offset} \in$ [-20, 20] voxels, $y_{offset} \in$ [-20, 20] voxels, $\theta \in$ [10, 30] deg, $\sigma_{blur}$ = [0, 0.6] \vspace{0.5em}} \\
  \hline
  \parbox[c][][t]{2cm}{Banding (e.g. due to magnetic field inhomogeneity)}& 
  \parbox[c][][t]{3cm}{Noise in a random spatial band  + Blur}& 
  \parbox[c][][t]{7cm}{\vspace{0.5em} \textbf{Class 1:} \\ Same set of parameters as used in noise. \\ \textbf{Class 2:} \\ Same set of parameters as used in noise. \vspace{0.5em}} \\
  \hline
  \parbox[c][][t]{2cm}{Motion (e.g. due to patient movement during scan)}& 
  \parbox[c][][t]{3cm}{Motion artefacts + Blur}& 
  \parbox[c][][t]{7cm}{\vspace{0.5em} \textbf{Class 1:} \\ $\theta \in$ [-10, 10] deg, motion offset $\in$ [0,3] mm, No. of transforms = 2, $\sigma_{blur}$ = [0, 0.6] \\ \textbf{Class 2:} \\  $\theta \in$ [-20, 20] deg, motion offset $\in$ [0,7] mm, No. of transforms = 4, $\sigma_{blur}$ = [0, 0.6]  \vspace{0.5em}} \\
  \hline
  \parbox[c][][t]{2cm}{Contrast (e.g. due to bias field)}& 
  \parbox[c][][t]{3cm}{Bias field heterogeneity + Gamma correction + Blur}& 
  \parbox[c][][t]{7cm}{\vspace{0.5em} \textbf{Class 1:} \\ Coefficients of polynomial for bias field $\in$ [0, 0.3], order of polynomial = 3, $\gamma \in$ [0, 0.3], $\sigma_{blur}$ = [0, 0.6] \\ \textbf{Class 2:} \\ Coefficients of polynomial for bias field $\in$ [0, 0.6], order of polynomial = 5, $\gamma \in$ [0.3, 0.6], $\sigma_{blur}$ = [0, 0.6]  \vspace{0.5em}} \\
  \hline
  \parbox[c][][t]{2cm}{Distortion (e.g. from gradient non-linearities and poor shimming)}& 
  \parbox[c][][t]{3cm}{Elastic deformation + Blur}& 
  \parbox[c][][t]{7cm}{\vspace{0.5em} \textbf{Class 1:} \\ Noise parameters are same as those for noise. No. of control points = 7, max. displacement = 9 voxels, Interpolation method = `bspline', $\sigma_{blur}$ = [0, 0.6] \\ \textbf{Class 2:} \\ Noise parameters are same as those for noise. No. of control points = 12, max. displacement = 12 voxels, Interpolation method = `bspline', $\sigma_{blur}$ = [0, 0.6] \vspace{0.5em}} \\
  \hline
%   \multicolumn{6}{c}{\parbox[c][][t]{11.5cm}{\centering }}\\
  \end{tabular}}
\vspace{-2.5em}
\label{tab:transforms}
\end{table*}
\begin{figure}[h!]
\centering
\includegraphics[width=11cm]{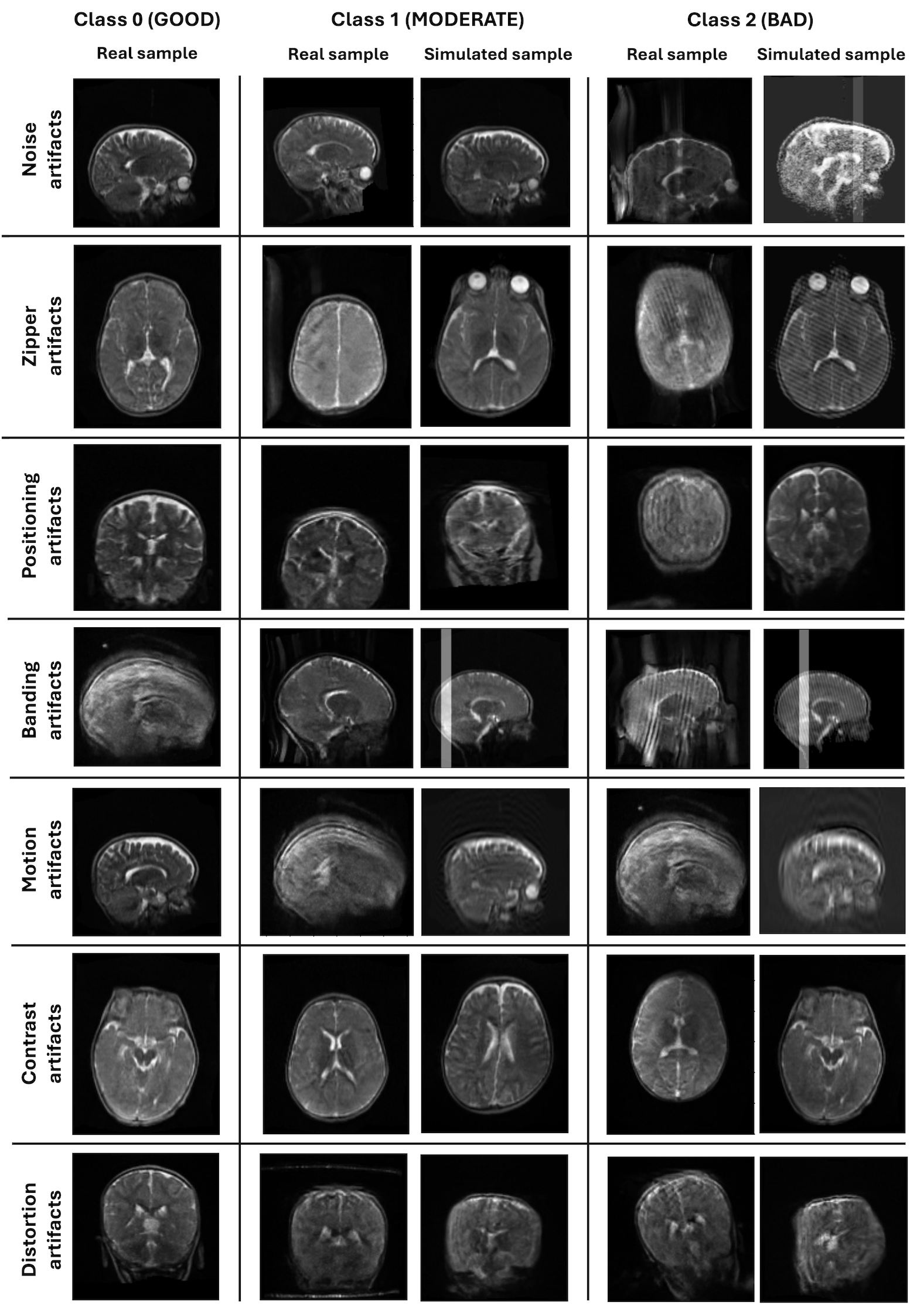}
\vspace{-1em}
\caption{Simulation of artefacts for various artefact domains for training the quality assessment models.}
\label{fig:sim}
\vspace{-2em}
\end{figure}
\subsubsection{Appearance-informed transformations for simulating artefacts.}
\label{sssec:transform}
We used various appearance-based transformation as specified in Table~\ref{tab:transforms} and a few examples are shown in Fig.~\ref{fig:sim} for various artefact domains. We used Torch IO library \cite{torchio2021} for simulating the artefacts using parameters mentioned in the table. Note that these transformations (specified in Table~\ref{tab:transforms}) were used for quality assessment (task 1) alone, since they were specific to the artefacts observed in the data.
\subsubsection{Automated quality assessment using deep learning.}
For classification of images based on their quality, we compared the following architectures:
\begin{itemize}
    \item Multi-headed decoder with single shared encoder \cite{hafiz2023se}
    \item DenseNet architecture \cite{huang2017densely}
\end{itemize}

Of the two architectures, this study is the first ever to experiment with multi-headed decoder framework, to the best of our knowledge, while DenseNet has been already used for similar tasks \cite{Ahmad2023}. The multi-headed decoder was chosen due to its capability to learn common attributes of data due to a shared encoder, which helps in learning perturbations better.

\begin{figure}[h!]
\centering
\includegraphics[width=\textwidth]{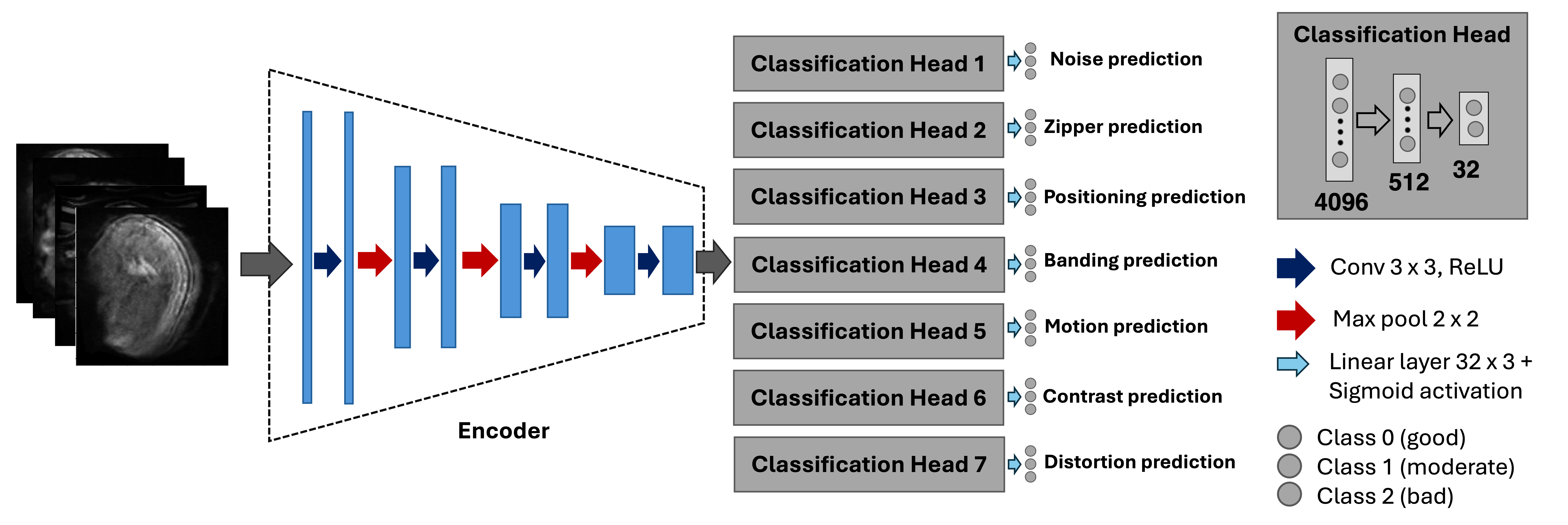}
\vspace{-1em}
\caption{Architecture of the multi-head decoder model for quality assessment.}
\label{fig:mhd}
% \vspace{-1em}
\end{figure}

\noindent\textbf{Multi-head decoder model for quality assessment. }
Figure~\ref{fig:mhd} shows the network architecture of the multi-head decoder model. The architecture consists of an encoder that is shared across multiple heads of the multiple decoder, that essentially extracts features that are different from high quality images without any artefacts. We used the encoder of a 4-layer deep UNet \cite{Ronneberger2015} model, and connected the decoders to the bottleneck of the UNet. Each decoder consisted of 3 fully connected layers (with 4096, 512 and 32 nodes respectively), followed by the output Sigmoid layer with 3 nodes (for classes 0, 1 and 2). The model was trained using an AdamW optimiser with a learning rate of $1\times10^{-5}$ and a patience value of 5. We used a batch size of 8, and train:validation split of 80:20. For standard data augmentations (note that these are separate from the transformations in section~\ref{sssec:transform}), we used and MONAI transforms for inflating the training data. We used focal loss (eqn.~\ref{eqn:focal}) \cite{lin2017focal} for training. The focal loss is given by:
% \vspace{-1em}
\begin{equation}
FL(p_t) = -\alpha_t(1-p_t)^\gamma log(p_t)
% , \quad p_t = 
% \begin{cases}
%     p, & \text{if} ~ y = 1 \\
%     (1 - p), & \text{otherwise}
% \end{cases} 
\label{eqn:focal}    
\end{equation}
where $\alpha$ and $\gamma$ are weighing and focusing parameters with values of 0.25 and 2 respectively (chosen empirically), and $p_t \in \{0,1\}$ is the predicted probability. During inference, we applied the trained encoder and individual decoders on each test instances and obtained the predictions for 7 artefact domains and used max-voting for obtaining final artefact prediction.

\noindent\textbf{DenseNet model architecture for quality assessment. }
We also trained a DenseNet \cite{huang2017densely} model using cross-entropy loss for training, individually for each artefact domain. We used the same parameters as used for training the multi-head decoder model.

% \noindent\textbf{Implementation details: }
\subsection{Segmentation of hippocampus}
Preliminary results from a 3D UNet indicated that the low contrast images were causing challenges for the segmentation of the hippocampus, leading to a large degree of under-segmentation. Therefore, we considered a range of different methods based on different paradigms:
\begin{itemize}
    \item Out-of-the-box Segmentation Approach (FSL FIRST \cite{patenaude2011})
    \item Linear Registration of an atlas
    \item 3d UNet
    \item 3d UNet + Prior
\end{itemize}

\noindent\textbf{Out-of-the-box (OOB) Segmentation Approach (FSL FIRST).} The majority of OOB segmentation tools for the hippocampus have been developed for adults and 1.5/3T MR images. Therefore, their expected performance on our images is unknown. We compared results on a subset of the training dataset for the following methods: FSL FIRST \cite{patenaude2011}, Freesurfer \cite{fischl2002}, SynthSeg \cite{billot2023} and HippoDeep \cite{thyreau2018}. Only FSL FIRST was able to reliably locate the hippocampus in the images and so is used as the OOB approach. 
\\

\noindent\textbf{Linear Registration.} For the linear registration approach we aimed to create a study specific hippocampus atlas, that could be then propagated to the individual subjects. We used FSL FLIRT \cite{jenkinson2002} to register all subjects to the first training subject (id: 0001), and then propagated all labels to this subject space and averaged them to produce a study specific hippocampus atlas. Non-linear registration was not used as due to the lack of contrast the algorithm struggled to converge and led to poor registration results. The atlas was then propagated back to the individual subject spaces for the validation samples, and thresholded at 0.1 (chosen empirically from held-out samples) to produce the binary segmentation mask. The average mask can be seen in Fig. \ref{fig:examplesubjs}C. 
\\

\begin{figure}[t]
\centering
\includegraphics[width=0.6\textwidth]{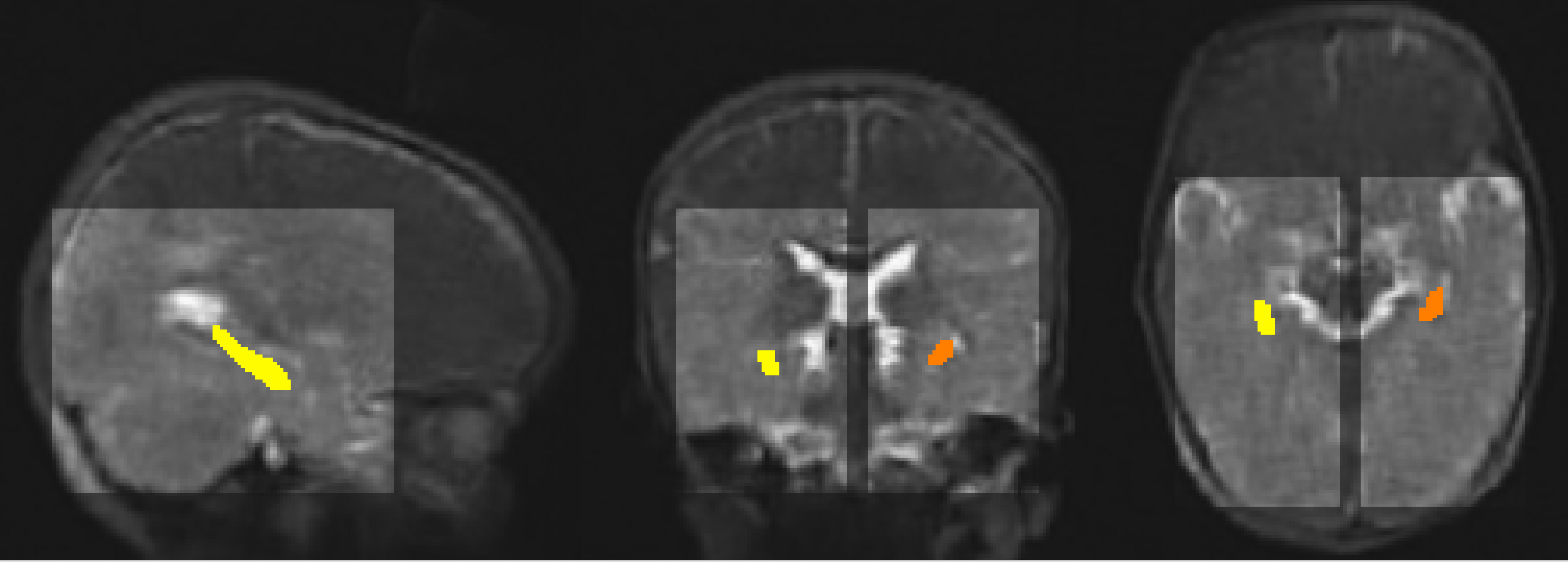}
\vspace{-1em}
\caption{Selected ROI from T2 }
\label{fig:ROI}
% \vspace{-1em}
\end{figure}

\noindent\textbf{3D UNet.} We considered a Vanilla 3D UNet \cite{Ronneberger2015} (64 features at the first level and 4 pooling layers), and trained with a Dice Loss function. Given the images were rigidly aligned, we selected an ROI from the training images centred around each hippocampus separately, as shown in Fig \ref{fig:ROI}. The images were split at the subject level into training and validation sets, and the model was trained using an AdamW optimiser with learning rate $1\times10^{-4}$ and a patience value of 15. The problem was treated as a binary segmentation task with both hippocampi being represented by the same label value, and standard augmentation was applied throughout training.
\\

\noindent\textbf{3D UNet + Prior.} On the held-out validation set, the 3D UNet was seen to have undersegemented the hippocampi, and so a prior was added to encourage the network to create larger segmentations. Specifically, an estimate of the amount of the ROI which should be hippocampus was calculated from the prior produced during the linear registration approach. This was binarised and then the ratio calculated. The model was then trained, inspired by \cite{Bateson2022}, using a loss function that directly aims to match the ratio of background to hippocampus:
\begin{equation}
    L_{total} = L_{dice}(\bm{Y}, \hat{\bm{Y}}) + \lambda KL(\tau, \hat{\tau})
\end{equation}
where $\bm{Y}$ is the true segmentation mask, $\bm{\hat{Y}}$ is the predicted segmentation mask, KL is the KL divergence between the true tissue ratio $\tau$, and the predicted tissue ratio $\hat{\tau}$. $\lambda$ is the weighting factor between the two loss functions, set to 5 empirically.  
\\

\noindent\textbf{Datasets used.} 
The data used were from the LISA Challenge \cite{LISA2024} consisting of low-field MRI dataset of over 300 paediatric T2 scans, acquired using a 0.064T MR Scanner with a sequence type of spin echo, TR 1.5s, TE 5ms and TI 400ms. The images were split (at the subject level) into training and validation sets with a split of 80:20\%. Quality assessment scores (values of 0, 1 and 2) were available across seven artefact domains (noise, zipper, positioning, banding, motion, contrast, distortion) in the CSV file format. Results are reported (section~\ref{ssec:qa_results}) on the 14 validation samples available as part of the Challenge. %The data used was a part of the LISA Challenge \cite{LISA2024}.
For segmentation task (task 2), the models were trained using the 79 manually labelled T2 MR images with manual labels. No preprocessing was performed other than those detailed in methods. Results are reported on the 12 validation samples, preprocessed identically to the training data where appropriate.

\section{Results and discussion}
\subsection{Quality Assessment Results}
\label{ssec:qa_results}
\begin{table}[h!]
\centering
\caption{Quality assessment results averaged across the dataset for various evaluation metrics. MHD: Multi-head decoder. WOA: without using augmentations, WA: With augmentations.}
\label{tab:qc_results}
\begin{tabular}{l|c|c|c|c}
\hline
Metric & MHD (WOA) & MHD (WA) & DenseNet (WOA) & DenseNet (WA)\\ \hline
$F1_{micro}$  & 0.357 & 0.741 & 0.735 & \textbf{0.823}\\
$F1_{macro}$ & 0.285 & 0.450 & 0.444 & \textbf{0.510}\\ 
\hline
$F1_{weighted}$ & 0.450 & 0.767 & 0.762 & \textbf{0.818}\\
\hline
$F2_{micro}$ & 0.357 & 0.741 & 0.735 & \textbf{0.823}\\
$F2_{macro}$ & 0.288 & 0.472 & 0.468 & \textbf{0.514}\\
\hline
$F2_{weighted}$ & 0.372 & 0.748 & 0.742 & \textbf{0.819}\\
\hline 
$Precision_{micro}$ & 0.357 & 0.741 & 0.737 & \textbf{0.823}\\
$Precision_{macro}$ & 0.421 & 0.510 &  0.446 & \textbf{0.554}\\
\hline
$Precision_{weighted}$ & 0.770 & 0.819 & 0.817 & \textbf{0.834}\\ 
\hline
$Recall_{micro}$ & 0.357 & 0.741 & 0.735 & \textbf{0.823}\\
$Recall_{macro}$ & 0.443 & \textbf{0.532} & 0.529 & 0.526\\
\hline 
$Recall_{weighted}$ & 0.357 & 0.741 & 0.735 & \textbf{0.823} \\
\hline
$Accuracy_{micro}$ & 0.357 & 0.741 & 0.735 & \textbf{0.823}\\
$Accuracy_{macro}$ & 0.357 & 0.741  & 0.735 & \textbf{0.823}\\
\hline
$Accuracy_{weighted}$ & 0.357 & 0.741 & 0.735 & \textbf{0.823}\\
\hline
\end{tabular}%
\end{table}
The results of comparison of multi-head decoder with DenseNet model are shown in Table~\ref{tab:qc_results}. As mentioned earlier, the main challenge in this task is the heavy class imbalance between class 0 and others (especially class 2, which are very low), hence leading the model to be biased towards class 0 and the anatomical structures are not being clearly distinguishable leading to poor predictive quality. However, it can be seen that the performance improved with the addition of simulated data using transformations described table~\ref{tab:transforms}. Between the architectures, DenseNet provided much better performance ($Accuracy_{weighted}: 0.823$), when compared to the multi-head decoder model ($Accuracy_{weighted}: 0.741$), likely   due to the ability of the former to extract complex spatial and contextual features that explain the subtle changes in the brain structure.
% \begin{figure}[h!]
% \centering
% \includegraphics[width=0.4\textwidth]{manual_volumes.png}
% \vspace{-1em}
% \caption{Comparison of left and right hippocampal volumes from the training dataset manual labels.}
% \label{fig:manualvolumes}
% \vspace{-1em}
% \end{figure}
\subsection{Segmentation Results}
The results comparing the different segmentation approaches are reported in Table \ref{tab:seg_results}. It can be seen that the OOB produced poor quality segmentations for the low-field data, as expected due to the large domain shift between this data and the data FSL FIRST was trained on. The vanilla UNet produced the better results, although far below the performance that would be expected for hippocampal segmentation on adult high field data. The addition of the prior reduced the segmentation quality slightly but the performance was very asymmetric (left 0.52±0.28, right 0.59±0.19), likely due the significantly different sizes of the manual masks for each hemisphere (left volume: 1160 $\pm$ 308, right volume: 1225 $\pm$ 338, paired ttest p=0.0008)). Registration of the average hippocampal volume provided the best results across the metrics apart from relative volume error, probably indicating that the atlas over estimates the size of the hippocampus. This demonstrates the lack of signal available in the images, as even very simple DL-based approaches would be expected to out-perform registration based approaches.  

\begin{figure}[t]
\centering
\includegraphics[width=0.8\textwidth]{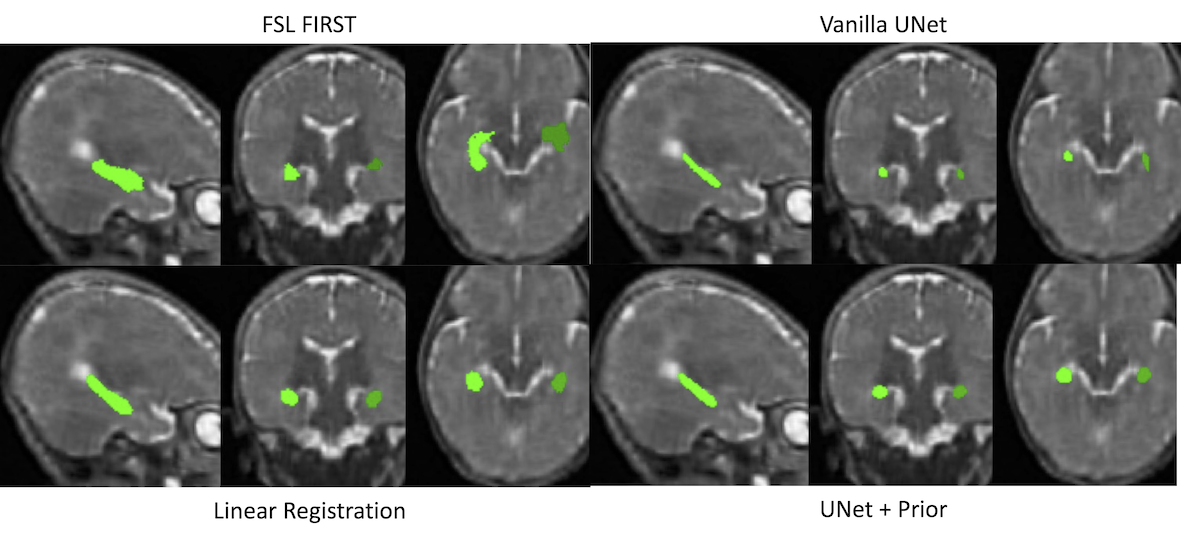}
\vspace{-1em}
\caption{Validation example segmentation results comparing the result from the four methods. }
\label{fig:validation}
\vspace{-1em}
\end{figure}

\begin{table}[h!]
\centering
\caption{Segmentation results averaged across left and right hippocampi for the five different metrics considered.}
\label{tab:seg_results}
\resizebox{0.95\columnwidth}{!}{%
\begin{tabular}{l|ccccc}
Method & Dice & \multicolumn{1}{c}{HD} & \multicolumn{1}{c}{HD95} & \multicolumn{1}{c}{ASSD} & \multicolumn{1}{c}{RVE} \\ \hline
FSL FIRST & 0.20±0.22 & 53.26±81.28 & 50.30±82.44 & 8.34±10.17 & 0.69±0.47 \\
Linear Registration & \textbf{0.61±0.15} & \textbf{7.32±8.03} & \textbf{3.12±0.89} &  \textbf{1.03±0.43} & 0.37±0.20\\
3D UNet & 0.58±0.18 & 8.94±8.22 & 4.44±2.17 & 1.24±0.94 & \textbf{0.22±0.14} \\
3D UNet + Prior & 0.56±0.22 & 11.50±12.51 & 6.74±8.78 & 2.46±4.00 & 11.05±37.56\\
\end{tabular}%
}
\end{table}

Figure \ref{fig:validation} shows an example segmentation from the validation data (no manual label available), comparing the four approaches. It can clearly be seen that FIRST over segments the hippocampus whereas the 3D UNet produces the lowest volume segmentations. The addition of the prior clearly increases the size of the segmented region. The localisation of the hippocampus is consistent across the linear registration, Vanilla 3D UNet and 3D UNet + prior approaches. 
\\
\begin{figure}[h!]
\centering
\includegraphics[width=0.75\textwidth]{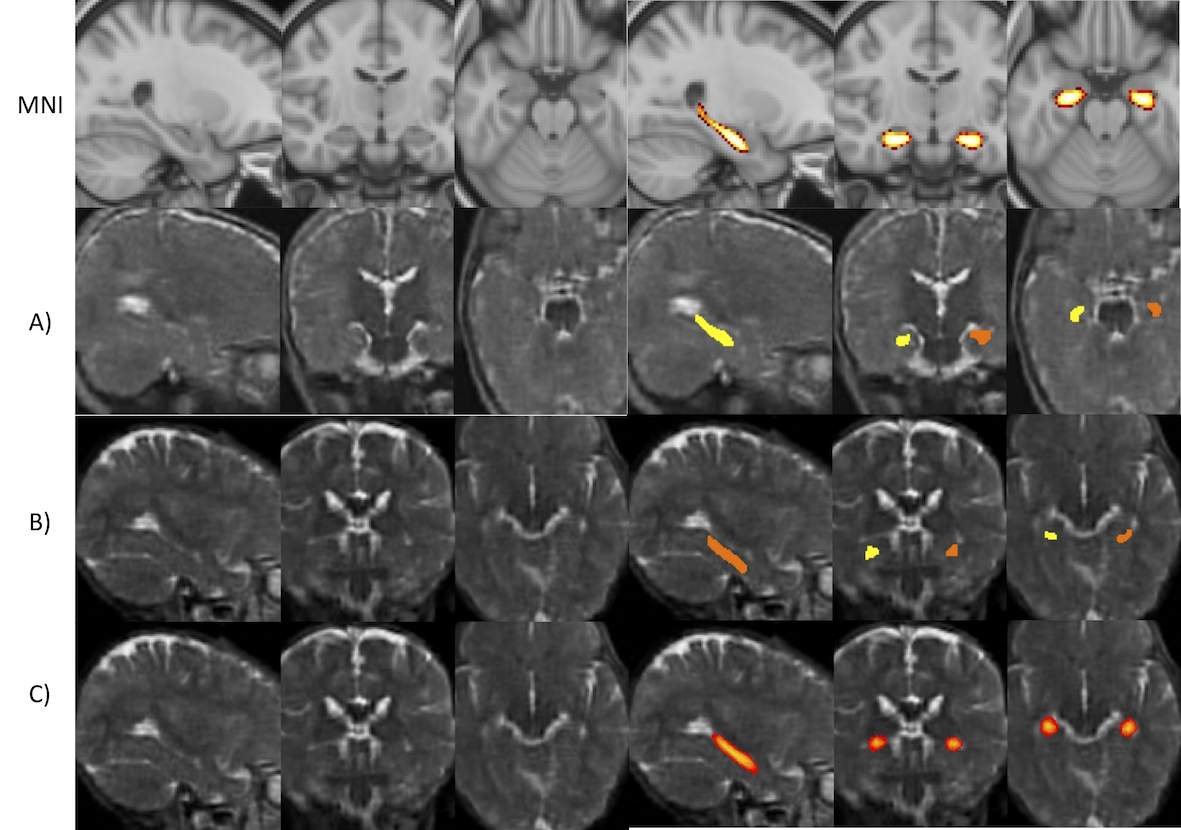}
\vspace{-1em}
\caption{Example hippocampal segmentations. Row 1: MNI T1 2mm template with the Harvard-Oxford probabilistic atlas, thresholded at 0.5. A): Example A showing a sample the 3D UNet repeatedly performed well (DSC $>$ 0.6) on with the tail of the atlas touching the ventricle. B) Example B showing a sample where the 3D UNet performed poorly  (DSC $>$ 0.1), where the localisation of the hippocampus in the target appears different to the MNI template and sample A, with the label not reaching the ventricles. C) Example B with the average hippocampus mask registered to it, showing that the average hippocampus mask is located disjoint to the manual segmentation mask.}
\label{fig:examplesubjs}
\vspace{-1em}
\end{figure}

Training and evaluation of the DL-based models were affected by the quality of the manual masks: unsurprisingly given the low contrast and relatively poor image quality, the quality of the training masks varied, biasing results. For instance, some samples appear to have labels which mislocalise the hippocampus (Fig. \ref{fig:examplesubjs}) where it can be seen that the example B's mask appears to be shifted lower than the expected location when considering the other examples and the average hippocampus mask registered to that subject. 

\section{Conclusions}
\label{sec:conc}
In this work, we provided preliminary analysis on low-field paediatric brain for two tasks: quality assessment and hippocampal segmentation, by comparing multiple approaches. Our results for quality assessment show that simulated artefacts helped to counteract the class imbalance while a densely connected architecture provided 10\% increase in accuracy. For hippocampal segmentation, we obtained the best results with registration of the study specific average, surpassing out-of-box methods which originally developed mainly for adult brains and deep learning approaches. As future work, more complicated DL models (e.g. transformer-based networks) would be implemented using self-supervised learning approach for increasing the performance. However, a vast amount of improvement would be required before the segmentation outputs can be used for automated granular analyses of paediatric brains. The Python implementation of our code is publicly available at \url{https://github.com/v-sundaresan/LISA2024_QA}.

\begin{credits}
\subsubsection{\ackname} 
\label{sec:ack}
This work was supported by DBT/Wellcome Trust India Alliance Fellowship [grant number IA/E/22/1/506763]. This work was also supported in part by Start-up Research Grant [grant number SRG/2023/001406] from the Science and Engineering Research Board, India and in part by Siemens Healthineers-CDS Collaborative Laboratory of Artificial Intelligence in Precision Medicine. VS is also supported by Pratiksha Trust, Bangalore, India [grant number FG/PTCH-23-1004] and the Seed Research Grant [grant number IE/RERE-22-0583] from the Indian Institute of Science, Bangalore, India.

\subsubsection{\discintname}
The authors have no competing interests to declare that are relevant to the content of this article.

\end{credits}

\bibliographystyle{splncs04}
\bibliography{references}

\end{document}